\documentclass[prl,twocolumn,superscriptaddress,showpacs,amssymb,amsmath,amsfonts,aps]{revtex4}

\usepackage{amsmath}
\usepackage{graphicx}
\usepackage{verbatim}
\usepackage{comment}

\begin{document}


\title{A Precise Measurement of the Neutron Magnetic Form Factor $G_M^n$ \\ in the Few-GeV$^2$ Region}




\newcommand*{\CMU}{Carnegie Mellon University, Pittsburgh, Pennsylvania 15213}
\affiliation{\CMU}
\newcommand*{\ODU}{Old Dominion University, Norfolk, Virginia 23529}
\affiliation{\ODU}
\newcommand*{\HU}{Hampton University, Hampton, Virginia 23668}
\affiliation{\HU}
\newcommand*{\MAINZ}{Institut f\"ur Kernphysik, Johannes Gutenberg-Universit¨at, 55099 Mainz, Germany}
\affiliation{\MAINZ}
\newcommand*{\UTFSM}{Universidad T\'{e}cnica Federico Santa Mar\'{i}a, Casilla 110-V Valpara\'{i}so, Chile}
\affiliation{\UTFSM}
\newcommand*{\URICH}{University of Richmond, Richmond, Virginia 23173}
\affiliation{\URICH}
\newcommand*{\JLAB}{Thomas Jefferson National Accelerator Facility, Newport News, Virginia 23606}
\affiliation{\JLAB}
\newcommand*{\OSUL}{Ohio State University, Lima, Ohio, 45804}
\affiliation{\OSUL}
\newcommand*{\UNIONC}{Union College, Schenectady, NY 12308}
\affiliation{\UNIONC}
\newcommand*{\ANL}{Argonne National Laboratory, Argonne, Illinois 60439}
\affiliation{\ANL}
\newcommand*{\ASU}{Arizona State University, Tempe, Arizona 85287-1504}
\affiliation{\ASU}
\newcommand*{\UCLA}{University of California at Los Angeles, Los Angeles, California  90095-1547}
\affiliation{\UCLA}
\newcommand*{\CSU}{California State University, Dominguez Hills, Carson, CA 90747}
\affiliation{\CSU}
\newcommand*{\CUA}{Catholic University of America, Washington, D.C. 20064}
\affiliation{\CUA}
\newcommand*{\SACLAY}{CEA-Saclay, Service de Physique Nucl\'eaire, 91191 Gif-sur-Yvette, France}
\affiliation{\SACLAY}
\newcommand*{\CNU}{Christopher Newport University, Newport News, Virginia 23606}
\affiliation{\CNU}
\newcommand*{\UCONN}{University of Connecticut, Storrs, Connecticut 06269}
\affiliation{\UCONN}
\newcommand*{\ECOSSEE}{Edinburgh University, Edinburgh EH9 3JZ, United Kingdom}
\affiliation{\ECOSSEE}
\newcommand*{\FU}{Fairfield University, Fairfield CT 06824}
\affiliation{\FU}
\newcommand*{\FIU}{Florida International University, Miami, Florida 33199}
\affiliation{\FIU}
\newcommand*{\FSU}{Florida State University, Tallahassee, Florida 32306}
\affiliation{\FSU}
\newcommand*{\GWU}{The George Washington University, Washington, DC 20052}
\affiliation{\GWU}
\newcommand*{\ECOSSEG}{University of Glasgow, Glasgow G12 8QQ, United Kingdom}
\affiliation{\ECOSSEG}
\newcommand*{\ISU}{Idaho State University, Pocatello, Idaho 83209}
\affiliation{\ISU}
\newcommand*{\INFNFR}{INFN, Laboratori Nazionali di Frascati, 00044 Frascati, Italy}
\affiliation{\INFNFR}
\newcommand*{\INFNGE}{INFN, Sezione di Genova, 16146 Genova, Italy}
\affiliation{\INFNGE}
\newcommand*{\ORSAY}{Institut de Physique Nucleaire ORSAY, Orsay, France}
\affiliation{\ORSAY}
\newcommand*{\ITEP}{Institute of Theoretical and Experimental Physics, Moscow, 117259, Russia}
\affiliation{\ITEP}
\newcommand*{\JMU}{James Madison University, Harrisonburg, Virginia 22807}
\affiliation{\JMU}
\newcommand*{\KYUNGPOOK}{Kyungpook National University, Daegu 702-701, Republic of Korea}
\affiliation{\KYUNGPOOK}
\newcommand*{\MIT}{Massachusetts Institute of Technology, Cambridge, Massachusetts  02139-4307}
\affiliation{\MIT}
\newcommand*{\UMASS}{University of Massachusetts, Amherst, Massachusetts  01003}
\affiliation{\UMASS}
\newcommand*{\UNH}{University of New Hampshire, Durham, New Hampshire 03824-3568}
\affiliation{\UNH}
\newcommand*{\NSU}{Norfolk State University, Norfolk, Virginia 23504}
\affiliation{\NSU}
\newcommand*{\OHIOU}{Ohio University, Athens, Ohio  45701}
\affiliation{\OHIOU}
\newcommand*{\PITT}{University of Pittsburgh, Pittsburgh, Pennsylvania 15260}
\affiliation{\PITT}
\newcommand*{\RPI}{Rensselaer Polytechnic Institute, Troy, New York 12180-3590}
\affiliation{\RPI}
\newcommand*{\RICE}{Rice University, Houston, Texas 77005-1892}
\affiliation{\RICE}
\newcommand*{\MOSCOW}{Skobeltsyn Nuclear Physics Institute, 119899 Moscow, Russia}
\affiliation{\MOSCOW}
\newcommand*{\SCAROLINA}{University of South Carolina, Columbia, South Carolina 29208}
\affiliation{\SCAROLINA}
\newcommand*{\VT}{Virginia Polytechnic Institute and State University, Blacksburg, Virginia   24061-0435}
\affiliation{\VT}
\newcommand*{\VIRGINIA}{University of Virginia, Charlottesville, Virginia 22901}
\affiliation{\VIRGINIA}
\newcommand*{\WM}{College of William and Mary, Williamsburg, Virginia 23187-8795}
\affiliation{\WM}
\newcommand*{\YEREVAN}{Yerevan Physics Institute, 375036 Yerevan, Armenia}
\affiliation{\YEREVAN}
\newcommand*{\NOWOHIOU}{ Ohio University, Athens, Ohio  45701}
\newcommand*{\NOWINDSTRA}{ Systems Planning and Analysis, Alexandria, Virginia 22311}
\newcommand*{\INDSTRA}{ Systems Planning and Analysis, Alexandria, Virginia 22311}
\newcommand*{\NOWGWU}{ The George Washington University, Washington, DC 20052}
\newcommand*{\NOWCNU}{ Christopher Newport University, Newport News, Virginia 23606}
\newcommand*{\NOWCUA}{ Catholic University of America, Washington, D.C. 20064}
\newcommand*{\NOWECOSSEE}{ Edinburgh University, Edinburgh EH9 3JZ, United Kingdom}
\newcommand*{\NOWLANL}{ Los Alamos National Laboratory, New Mexico, NM}
\newcommand*{\LANL}{ Los Alamos National Laboratory, New Mexico, NM}
\newcommand*{\NOWKYUNGPOOK}{ Kyungpook National University, Daegu 702-701, Republic of Korea}
\newcommand*{\NOWJLAB}{ Thomas Jefferson National Accelerator Facility, Newport News, Virginia 23606}
\newcommand*{\MASON}{ George Mason University, Fairfax, Virginia 22030}

\author {J.~Lachniet} 
\affiliation{\CMU}
\affiliation{\ODU}

\author{A.~Afanasev}
\affiliation{\HU}

\author{H.~Arenh\"ovel}
\affiliation{\MAINZ}

\author {W.K.~Brooks} 
\affiliation{\UTFSM}

\author {G.P.~Gilfoyle} 
\affiliation{\URICH}

\author {D.~Higinbotham} 
\affiliation{\JLAB}

\author {S.~Jeschonnek}
\affiliation{\OSUL}

\author {B.~Quinn} 
\affiliation{\CMU}

\author {M.F.~Vineyard} 
\affiliation{\UNIONC}

\author {G.~Adams} 
\affiliation{\RPI}
% ...
\collaboration{The CLAS Collaboration}
     \noaffiliation


\date{\today}

\begin{abstract}
The neutron elastic magnetic form factor was extracted from
quasielastic electron scattering on deuterium over the range  $Q^2 =1.0~\rm{GeV^2} - 4.8~\rm{GeV^2}$
with the CLAS detector at Jefferson Lab. 
High precision was achieved with a ratio technique and 
a simultaneous in-situ calibration of the neutron
detection efficiency. Neutrons were detected 
with electromagnetic calorimeters and time-of-flight scintillators at two beam energies.
The dipole parameterization gives a good description of the data.
\end{abstract}


\maketitle


The elastic electromagnetic form factors of the proton and neutron are fundamental
quantities related to their spatial charge and current distributions.
The dominant
features of the larger form factors $G_M^p$, $G_E^p$, and $G_M^n$ were
established in the 1960's: the dipole form 
$G_{D} = (1+ Q^2/\Lambda)^{-2}$ where $\rm \Lambda = 0.71~\rm GeV^2$ gave a  
good description of these form factors ($G_M^p/\mu_p \approx G_M^n/\mu_n \approx G_E^p \approx G_{D}$)
within experimental uncertainties, corresponding 
(at least for $ Q^2 \ll 1 ~\rm GeV^2$ or large radii) to an exponential falloff in the spatial
densities of charge and magnetization. 
Recent Jefferson Lab results on the proton form factors show a dramatic departure from the dipole form
even at moderate $Q^2$ \cite{gayou2} while the neutron magnetic form factor $G_M^n$ falls 
below the dipole at high $Q^2$
($G_M^n/\mu_nG_D = 0.62\pm 0.15$ at $Q^2 = 10~{\rm GeV^2}$ \cite{rock2}).
Describing all these modern results with nucleon models and lattice calculations
has been a challenge \cite{miller1,Kroll2,guidal2,meissner2,thomas2}. 
Also, the elastic form factors are the zeroth moment of
the generalized parton distributions (GPDs) and thus, 
constrain GPD models \cite{Kroll2}. 
Last, we note that some models predict significant deviations from the dipole
for $Q^2 < 5~\rm GeV^2$ \cite{thomas2, guidal2}.

To distinguish among different models, high precision and large
$ Q^2$ coverage are important. 
At larger momentum transfer $G_M^n$ is known much more poorly than
the proton form factors \cite{Kees}.
In this Letter we report on a new measurement of $G_M^n$ in the range $ Q^2 = 1.0-4.8 ~ \rm GeV^2$ at Jefferson Lab.
The precision and coverage of these results eclipse  
the world's data in this $ Q^2$ range.
Systematic uncertainties were held to 2.5\% or less.

In the absence of a free neutron target, we measure the ratio $R$ of the cross sections for the
${\rm ^2H}(e,e^\prime n)p$  and ${\rm ^2H}(e,e^\prime p)n$ reactions 
in quasielastic (QE) scattering on deuterium.
A nucleon with most of the momentum from the scattered electron is detected in coincidence with
the final state electron.
The ratio $R$ is defined as 
$R={\frac{d\sigma}{d\Omega}}[{\rm ^2H}(e,e'n)_{QE}]/{\frac{d\sigma}{d\Omega}}[{\rm ^2H}(e,e'p)_{QE}]$ \cite{jl1,bartel2, anklin1,durand1} and 
\begin{align}
R  & = a(E,  Q^2,\theta_{pq}^{max}, W^2_{max}) \times \nonumber \\[2pt]
   &  \qquad\quad       \frac{\sigma_{Mott}\left ({\frac{(G_E^n)^2+\tau(G_M^n)^2}{1+\tau}} + 2\tau\tan^2{\frac{\theta}{2}}(G_M^n)^2 \right )}
             {{\frac{d\sigma}{d\Omega}}[{\rm  ^1H}(e,e')p]},\label{eq2}
\end{align}
\noindent where $E$ is the beam energy, $\sigma_{Mott}$ is the 
cross section for scattering off a scalar (spinless), point particle of unit charge, $\tau =  Q^2/4M^2$, $ M$ is the nucleon mass, and
$\theta$ is the electron scattering angle.
The factor $a(E, Q^2,\theta_{pq}^{max},W^2_{max})$ corrects for nuclear effects and depends on $E$ and cuts on
$\theta_{pq}^{max}$, the maximum angle between the nucleon direction and the three-momentum transfer $\vec q$,
and $W^2_{max}$, the square of the maximum value of the mass recoiling against the
electron assuming a stationary target.
We used the one-photon exchange approximation in the numerator of Eq.~\ref{eq2} to express the cross section in terms 
of the neutron form factors.
The right-hand side of Eq.~\ref{eq2} contains the desired $G_M^n$ along with the better-known proton cross section
and the neutron electric form factor ($G_E^n$), which is believed to be small over the $ Q^2$ range here.
For QE kinematics (within a cone $\theta^{max}_{pq}$ around $\vec q$) $G_M^n$ can be extracted  from Eq.~\ref{eq2} as a 
function of $ Q^2$ by relying on knowledge of the proton cross section ({\it i.e.}, the Arrington parameterization \cite{arringtonGMnFits2}), $G_E^n$, 
calculations of $a(E, Q^2,\theta_{pq}^{max},W^2_{max})$,
and measurements of $R$.
The ratio method is less vulnerable to nuclear structure ({\it e.g.}, choice of 
deuteron wave function, {\it etc.}) \cite{durand1} and experimental effects ({\it e.g.}, radiative
corrections, {\it etc.}).
The challenge here is to accurately measure the nucleon detection efficiencies.

The two reactions were measured in the CLAS detector \cite{nim} at the same time
and from the same target to reduce systematic uncertainties.
Two electron-beam energies were used, 2.6 GeV and 4.2 GeV.
CLAS consists of six independent magnetic spectrometers each instrumented with drift chambers \cite{CLASDC},
time-of-flight (TOF) scintillators covering polar angles $8^\circ < \theta <143^\circ$ \cite{CLASTOF}, a gas-filled threshold
Cherenkov counter (CC) \cite{CLASCC}, and a lead-scintillator sandwich-type electromagnetic calorimeter (EC) 
covering  $8^\circ < \theta <45^\circ$ \cite{CLASEC}.
CLAS was triggered on electrons by requiring a coincidence between CC and EC signals in one sector.
Neutrons were measured separately in the TOF and EC.
Protons were measured using the drift chambers and TOF systems.
A novel dual-cell target was used consisting of two collinear 
cells each 5-cm long  - one filled with $\rm ^1H$ and the other with $\rm ^2H$ - and separated by 4.7 cm.
The downstream cell was filled with liquid hydrogen for calibrations and efficiency measurements.
The upstream cell was filled with liquid deuterium for the ratio measurement.
The target was made of aluminum with 20-micron aluminum windows.
The CLAS vertex resolution of 2 mm enabled us to separate events from the 
different targets \cite{nim}.

We now describe the analysis.
Nucleons from quasielastic events tend to be ejected close to the direction of the 3-momentum transfer $\vec q$
while inelastically scattered nucleons are not \cite{durand1}.
We required the angle $\theta_{pq}$ between the nucleon 3-momentum and $\vec q$ to be small 
($\theta_{pq}^{max} = 2.5^\circ - 4.5^\circ$ across the $ Q^2$ range) and integrated over all azimuthal angles about $\vec q$.
Another cut,  $W^2 < W^2_{max} = 1.2~{\rm GeV^2}$ eliminated most inelastic events that survived the $\theta_{pq}^{max}$ cut.
Our simulations of the quasielastic \cite{jl1} and inelastic production \cite{GENEV} 
show the fraction of inelastic events surviving these cuts is less than 0.5\% of the total.
To measure $R$ accurately, the solid angles of CLAS for the ${\rm ^2H}(e,e^\prime n)_{QE}$ and ${\rm ^2H}(e,e^\prime p)_{QE}$ reactions have to be identical.
The nucleon solid angles were matched by first determining event-by-event the nucleon momentum from the
electron kinematics assuming quasielastic scattering.
The expected proton and neutron trajectories in CLAS were checked to see if both trajectories would lie within the
CLAS acceptance.
Only events where both nucleons were expected to strike CLAS were analyzed.

Once the event sample was selected, corrections for the detector efficiencies and other effects were applied.
Neutrons were measured in two CLAS scintillator-based detectors: the EC and
the TOF. The neutron detection efficiency (NDE) measurement was
performed using tagged neutrons from the ${\rm ^1H}(e,e'\pi^+)n$ reaction,
where the mass of the unobserved neutron was inferred from the measured electron and pion kinematics and matched with possible
hits in the neutron detector.
The value of the detection efficiency can vary with
time-dependent and rate-dependent quantities like
photomultiplier tube gain so it was measured
\emph{simultaneously} with the primary deuterium measurement. 
The measured neutron detection efficiency for each sector for the TOF and for nine subsections in each EC sector 
were fitted with polynomials at low neutron momenta 
and a constant at high
momenta. 
The EC efficiency typically reached a maximum value of $\approx \! 0.6$ while the maximum TOF efficiency 
was  $\approx \! 0.08$ \cite{jl1,CLASphysicsDB}.
The calibration target was also used to measure the proton detection efficiency using
elastic scattering $p(e,e^\prime p)$.
The kinematics of the scattered electron were used to predict the location of the elastically scattered proton
in CLAS and the event was searched for a proton at that location.

The calculation of the nuclear correction factor, $a(E, Q^2,\theta_{pq}^{max},W^2_{max})$, in Eq.~\ref{eq2}
is described in Ref.~\cite{sabine3b}.
The cross section was calculated using the Plane Wave Impulse Approximation (PWIA) for
$ Q^2 \ge 1.0~ \rm GeV^2$, the AV18 deuteron wave function \cite{av18b}, and Glauber theory for final-state interactions (FSI).
The correction is the ratio
of the full calculation to the PWIA without FSI.
The correction was averaged over the same $\theta_{pq}$ range used  in the analysis and 
was less than 0.1\% across the full $ Q^2$ range.

In our analysis we assumed QE kinematics and ignored the Fermi motion that can knock the ejected nucleon out of the acceptance.
To correct for this effect we simulated QE scattering from a fixed target nucleon and tested to see if it struck the
active area of CLAS (an `expected' event).
We then simulated the nucleon's internal motion (with the Hulthen distribution) and elastic scattering from this moving particle.
With the target momentum known (in simulation) we re-calculated the trajectory to see if it still struck 
CLAS and satisfied the $\theta_{pq}^{max}$ cut (an `actual' event).
The ratio of actual to expected events is the correction for that nucleon.
The ratio of these corrections for the neutron and the proton multiplies $R$.
The correction to $G_M^n$ is in the range $\approx 0.9 - 1.3$.

\begin{figure}
\begin{center}
\includegraphics[height=2.0in]{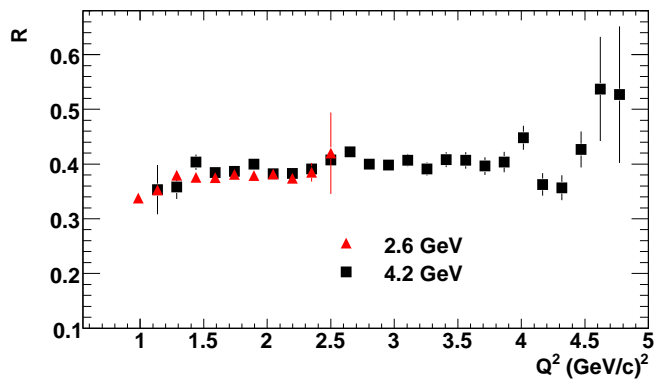}
\caption{\label{fig:Rplot1} (color online). Results for $R$ as a function of $ Q^2$ for two beam energies. 
Each set is a weighted sum of the TOF and EC neutron measurements. 
Only statistical uncertainties are shown. 
Numerical results are reported in the CLAS Physics Data Base \cite{CLASphysicsDB}.}
\end{center}
\end{figure}

We present our results for $R$ in Fig.~\ref{fig:Rplot1} for the two beam energies and for $Q^2 > 1~\rm GeV^2$ where we have
overlapping TOF and EC data.
The corrections described above have been included and only statistical uncertainties are shown.
For each beam energy we averaged the two neutron measurements (EC and TOF) weighted
by the statistical uncertainties.
Measurements of $R$ at the same $ Q^2$ but different beam energies
are not  expected to be the same because the kinematics are not the same (recall Eq.~\ref{eq2}).
The data cover the $ Q^2$ range with excellent statistical accuracy and with a large overlap between the two data sets.

A detailed study of each correction's contribution to the systematic uncertainty  has been performed \cite{jl1}.
Listed in Table \ref{table:errors} are the largest contributions along with the maximum (typical) value across the
full $ Q^2$ range.
\begin{table}[b!]
\begin{center}
\begin{tabular}{|l|c|l|c|} \hline
Quantity                                & $\delta G_M^n/G_M^n$     & Quantity                                & $\delta G_M^n/G_M^n$   \\ \hline
EC NDE                                  & $\quad <1.5\%$ (1\%)    & TOF NDE                  & $\quad <3.2\%$ (2\%)  \\ \hline
proton $\sigma$                         & $\quad <1.5\%$ (0.8\%)  &$G_E^n$                                 &  $\quad <0.7\%$ (0.5\%)  \\ \hline
Fermi loss                              & $\quad <0.9\%$ (0.5\%)  & $\theta_{pq}$ cut                       &  $\quad <1.0\%$ (0.3\%) \\ \hline
Remainder                               & $\quad <0.5\%$ (0.2\%)  & & \\ \hline
\end{tabular} 
\caption{Upper limits (typical values) of systematic errors.}\label{table:errors}
\end{center}
\end{table}
The largest contributions come from the parameterizations of the neutron detection efficiencies for the TOF
and EC systems.
To estimate the uncertainty associated with the NDE measurement,
the order of the polynomial and position of the edge of the constant region used to fit the data were varied
to determine the effect on $G_M^n$ as a function of $ Q^2$.
Uncertainties were in the range 0.5-3.2\%.

The extraction of $G_M^n$ depends on the other elastic form factors (see Eq.~\ref{eq2}) and their uncertainties
contribute to the uncertainty in $G_M^n$.
The proton cross section uncertainty was estimated using the difference
between parameterizations by Bosted and Arrington \cite{bosted, arringtonGMnFits2}.
The average difference was $<1$\% with a maximum of 1.5\%.
For $G_E^n$, the difference between the Galster parameterization
and a fit by Lomon was used \cite{galster2,lomon} with a maximum uncertainty of 0.7\%.
The upper limit of the $\theta_{pq}$ cut was varied by $\pm10$\%, changing $G_M^n$ by 
a maximum of about 1.0\% and by 0.3\% on average \cite{jl1}.
The uncertainty of the Fermi motion correction was calculated using two dramatically different momentum 
distributions of the deuteron: a flat one and the 
Hulthen distribution.
This correction to $G_M^n$ changes by $<1$\% between the two Fermi motion distributions.
The quadrature sum of the remaining, maximum systematic uncertainties was less than 0.5\% \cite{jl1}.
The final systematic uncertainty for the EC measurement was $<2.4$\% and for the TOF 
measurement it was $<3.6$\%.

\begin{figure}[hb] 
\begin{center}
\includegraphics[height=2.0in]{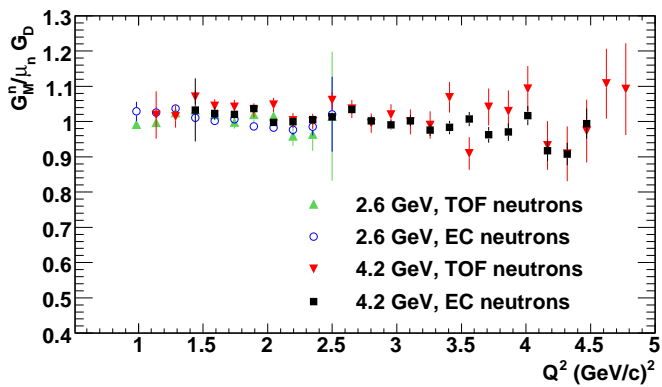}
\caption{\label{fig:gmn_detector_compare} (color online). Results for 
$G_M^n / (\mu_n G_{D}) $ as a function of $ Q^2$ for four 
different measurements (two beam energies). Only statistical uncertainties are shown.}
\end{center}
\end{figure}

The CLAS extraction of $G_M^n( Q^2)$ consists of overlapping measurements. 
The TOF scintillators cover the full angular range of CLAS, while the EC system covers
a subset of these angles, so $G_M^n$ can be obtained from two
independent measurements of the $e-n$ production. 
The experiment was performed with two beam energies with overlapping $ Q^2$ coverage so the detection of 
nucleons of a given $ Q^2$ occurs in two different regions of CLAS. 
Four measurements of
$G_M^n$ have been obtained from CLAS that
could have four semi-independent sets of systematic uncertainties. 
Shown in Fig.~\ref{fig:gmn_detector_compare} are the results for $G_M^n$ from the 
different measurements divided by $\mu_n G_{D}$ for normalization and to reduce the dominant $ Q^2$ dependence.
Only statistical uncertainties are shown.
Here the different measurements should agree because $G_M^n$ depends only on $ Q^2$.
The two measurements for each beam energy are consistent within the statistical uncertainties,
suggesting the systematic uncertainties are well-controlled and small.
The results in Fig.~\ref{fig:gmn_detector_compare} were then combined in a weighted average as a function of $ Q^2$.
The final systematic uncertainty varied from 1.7-2.5\% across the full data range.
The larger uncertainty on the parameterization of the TOF NDE (see Table 1) did not
push the total, weighted uncertainty above our goal of 3\%. There are more calorimeter data
due to its higher efficiency and the maximum EC uncertainty was 1.5\% \cite{jl1,CLASphysicsDB}.

\begin{figure}
\begin{center}
\includegraphics[height=2.4in]{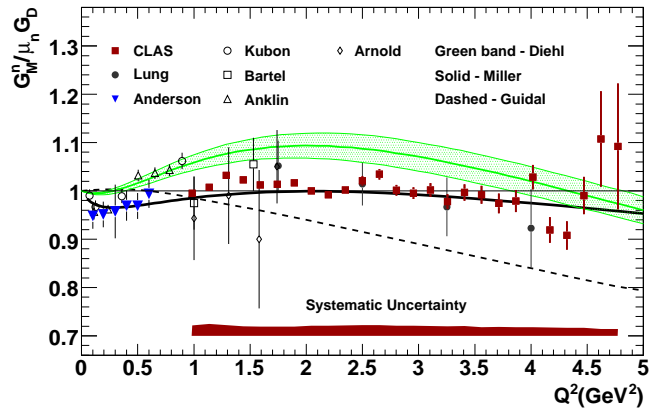}
\caption{\label{fig:final_results} (color online). Results for $G_M^n / (\mu_n G_{D}) $ from the CLAS measurement
are compared with a selection of previous data \cite{bartel2,kubon,lung2,anklin1,anderson,arnold2} and 
theoretical calculations \cite{Kroll2, guidal2, miller1}. 
Numerical results are reported in the CLAS Physics Data Base \cite{CLASphysicsDB}.}
\end{center}
\end{figure}

The final, combined results for $G_M^n$ are shown in Fig.~\ref{fig:final_results}
with a sample of existing data \cite{bartel2,kubon,lung2,anklin1,anderson,arnold2}. 
The uncertainties are statistical only. Systematic uncertainties are
represented by the band below the data.
A few features are noteworthy. 
First, the quality and coverage of the
data is a dramatic improvement of the world's data set. 
Second, our results are consistent with previous data, but with much smaller uncertainties.
Third, the dipole form is a good
representation here, which differs from parameterizations and some calculations
at higher $ Q^2$ where 
previous results for $G_M^n /(\mu_n G_{D})$ decrease with increasing $ Q^2$ \cite{Kees, guidal2,  thomas2}.
We note there appears to be an offset between the low-$Q^2$ end of our data and some earlier results \cite{kubon, anklin1}
that is about twice the uncertainty of the offset.
Last, any apparent fluctuations in our results ({\it e.g.} at $1.29~\rm GeV^2$) are not significant enough to draw any firm conclusions here.

The curves shown in Fig.~\ref{fig:final_results} are from
Diehl {\it et al.} \cite{Kroll2}, Guidal {\it et al.} \cite {guidal2}, and Miller \cite{miller1}
and are all constrained by the world's previous data.
In Diehl {\it et al.} the GPDs are parameterized and fitted to the experimental data (green band).
The curve reproduces some of the low-$Q^2$ data, but lies above our results.
Guidal {\it et al.} use a Regge parameterization of the GPDs to characterize the elastic nucleon form factors at
low momentum transfer and extend it to higher $ Q^2$ (dashed line). 
The curve reproduces the existing, higher $Q^2$ data (which fall well below the dipole in the range $Q^2 = 6-10~\rm GeV^2$),
but is not consistent with our results.
In Miller's calculation the nucleon is treated using light-front dynamics as a relativistic
system of three bound quarks and a surrounding pion cloud (solid curve).
The model gives a good description of much of the previous data 
even at high $Q^2$ and is consistent with our results.

The neutron magnetic form factor was measured in the range
$ Q^2 = 1.0-4.8~\rm GeV^2$ with the CLAS detector at Jefferson Lab using the ratio of $e-n$ to $e-p$ scattering.
Two incident beam energies were used and systematic uncertainties were $\le 2.5$\%.
Neutrons were measured with two independent systems: time-of-flight scintillators and electromagnetic calorimeters.
Detector efficiencies were measured simultaneously
with the production data using a dual-cell
target containing $^2\rm H$ and $^1\rm H$.
The data provide a significant improvement in precision and coverage in this $Q^2$ range and are surprisingly consistent with
the long-established dipole form.
The calculation by Miller is in good agreement with our results.

We acknowledge the outstanding efforts of the staff of the 
Accelerator and Physics Divisions at Jefferson Lab that made this experiment possible.
This work was supported in part by the Italian Istituto Nazionale di Fisica Nucleare, the 
 French Centre National de la Recherche Scientifique and Commissariat \`{a} l'Energie Atomique, the U.S. Department of Energy, the National 
Science Foundation, 
an Emmy Noether grant from the Deutsche Forschungsgemeinschaft, 
the U.K. Engineering and Physical Science Research Council,
the Chilean Fondo Nacional de Desarrollo Cientifico y Tecnol\'ogico,
and the Korean Science and Engineering Foundation.
Jefferson Science Associates operates the 
Thomas Jefferson National Accelerator Facility for the U.S. 
D.O.E. under contract DE-AC05-06OR23177.


\end{document}